\documentclass[12pt]{article}
\usepackage{epsfig,cite}
\usepackage {amsmath}
\usepackage{amssymb}
\usepackage{graphicx}
\usepackage{epstopdf}

\include{epsf}

\usepackage{times}

\def\beq{\begin{equation}}
\def\eeq{\end{equation}}
\def\bey{\begin{eqnarray}}
\def\eey{\end{eqnarray}}

\def\lsim{\mathrel{\raise.3ex\hbox{$<$\kern-.75em\lower1ex\hbox{$\sim$}}}}
\def\gsim{\mathrel{\raise.3ex\hbox{$>$\kern-.75em\lower1ex\hbox{$\sim$}}}}

\begin{document}

\begin{center}
UTTG-11-11
\hspace{0.3cm} TCC-013-11
\hspace{0.3cm} MCTP-11-30
\hspace{0.3cm} FERMILAB-PUB-11-385-A
\end{center}
\vskip 0.2in
\begin{center}
{\large{\bf  Gamma-Ray Constraints on the First Stars from Annihilation of Light WIMPs}}
\end{center}
\begin{center}
\vskip 0.2in
{\bf Pearl Sandick,}$^{1,2}$
{\bf Juerg Diemand,}$^3$
{\bf Katherine Freese,}$^{1,4}$
{\bf and Douglas Spolyar}$^{5,6}$

\vskip 0.1in

{\it $^1${Theory Group and Texas Cosmology Center, The University of Texas at Austin, TX 78712}\\}
{\it $^2${Department of Physics and Astronomy, University of Utah, Salt Lake City, UT 84112}\\}
{\it $^3${Institute for Theoretical Physics, University of Z\"urich, CH-8057, Switzerland}\\}
{\it $^4${Michigan Center for Theoretical Physics, University of Michigan, Ann Arbor, MI 48109}\\}
{\it $^5${Center for Particle Astrophysics, Fermi National Accelerator Laboratory, Batavia, IL  60510}\\}
{\it $^6${Department of Astronomy and Astrophysics, The University of Chicago, Chicago, IL 60637}\\}

\end{center}
\vskip 0.1in
\begin{abstract}

We calculate the limits on the fraction of viable dark matter minihalos in the early universe to host Population III.1 stars, surviving today as dark matter spikes in our Milky Way halo.  Motivated by potential hints of light dark matter from the DAMA and CoGeNT direct dark matter searches, we consider thermal relic WIMP dark matter with masses of 5, 10, and 20 GeV, and annihilation to $\mu^+\mu^-$, $\tau^+\tau^-$, and $q\bar{q}$.  From this brief study we conclude that, if dark matter is light, either the typical black hole size is $\lesssim 100M_\odot$ (i.e.~ there is no significant Dark Star phase), and/or dark matter annihilates primarily to $\mu^+\mu^-$ or other final states that result in low gamma-ray luminosity, and/or that an extremely small fraction of minihalos in the early universe that seem suitable to host the formation of the first stars actually did.

\end{abstract}



\section{Introduction}

The very first generation of stars, known as Population III.1, likely formed from the pristine gas at the
 centers of $\sim 10^6 M_\odot$ dark matter minihalos at $z \gtrsim 10$~\cite{HTL1996}.  The response of the dark matter in a minihalo to the formation of a compact baryonic object at its center is a contraction of the dark matter density profile in and around the object.  When the first stars, expected to be $\gtrsim 100 M_\odot$, ended their lives by collapsing to black holes\footnote{Stars in the mass range $\sim 140-260 M_\odot$ would have ended their lives as pair instability supernovae, leaving no remnants~\cite{hegerwoosley}.  We do not consider these objects here.}, each remnant remained surrounded by a region of enhanced dark matter density, which we call a {\it dark matter spike}. 
In fact, if the first stage of stellar evolution is a Dark Star phase, during which the star is powered by dark matter annihilations, the first stars would have grown to be even larger, leaving correspondingly larger black holes and surrounding dark matter spikes.

Many of these spikes, remnants of the formation of the first stars, may have fallen into our Galactic halo, constituting Milky Way dark matter substructure today. 
If the dark matter in the spikes is made of Weakly Interacting Massive Particles (WIMPs), typically these are their own antiparticles, and they annihilate with one
another inside the spikes surrounding the black holes.  This paper focuses on the gamma-ray flux from the annihilation, and compares it to data 
from the Fermi Gamma-Ray Space Telescope (FGST).  In previous work ~\cite{dmspikes, dmspikesConf}, hereafter SDFS, we considered WIMPS with masses in the range 100 GeV to 2 TeV and
used the FGST data~\cite{fgstEGB, fgstFSC} to constrain models of Population III.1 star formation and/or dark matter annihilation for these WIMP masses. Subsequently, the possibility that WIMP dark matter is non-thermal in origin was also investigated~\cite{sandickwatson}.
In this paper, we return to the thermal WIMP paradigm, but consider WIMP masses of 5, 10, and 20 GeV.  

The motivation for this study is 
the recent increase in attention to lower mass WIMPs due to possible hints of their discovery in direct and indirect detection experiments.  The DAMA and DAMA/LIBRA direct detection experiments~\cite{bernabei} have for ten years
found an annual modulation \cite{KTannmod} of their signal compatible with $\sim $10 GeV WIMPs, and now the CoGeNT\cite{CoGeNT} experiment claims a 2.8$\sigma$ annual modulation signal, expected to be compatible with that observed by DAMA~\cite{cogentAnnMod}. 
Since the two experiments are made of different detector materials (the relevant nucleus for DAMA is Na while for CoGeNT it is Ge), and the two experiments
are on different continents, the situation is interesting. 
Additionally, the CRESST-II experiment, using calcium tungstate (CaWO$_4$) as the target material, has recently announced an excess of events that also seems to be compatible with $\sim 10$ GeV WIMPs~\cite{cresst}.
Yet, the CDMS experiment, also made of germanium, claims to exclude the range of WIMP masses and cross sections indicated by the annual modulation signal~\cite{cdmsLDM}, as does the XENON100 experiment~\cite{xenon100}.  
Further potential evidence for light WIMP dark matter has also come from indirect dark matter searches in the form of an apparent excess in gamma-rays from the Galactic center region~\cite{HooperGoodenough}, while more recent gamma-ray studies of dwarf galaxies indicate that light WIMPs with the canonical thermal annihilation cross section are disfavored~\cite{GeringerSameth:2011iw}.
Certainly the experimental situation is unresolved,
but still it is interesting to consider the possibility that dark matter is made of lower mass WIMPs.  Here we examine the dark matter annihilation signal from light WIMP dark matter in spikes in our Galactic Halo and compare the expected signal to gamma-ray data from the Fermi satellite.

Several other groups have previously studied signatures of annihilation in dark matter overdensities, or spikes, around black holes~\cite{GS,ZS,bzs,bftz,tabp,bls} (see also the review in~\cite{dmspikes}).  Our work is similar in spirit and makes the additional step of direct comparison with FGST data.

\section{The First Stars and Their Dark Matter Spikes}
\label{sec:dmspikes}

Population III.1 stars likely formed from metal-free, molecular hydrogen-cooled gas at the center of dark matter minihalos.
 We use the parametrization of Ref.~\cite{Hcooling} for the minimum halo mass in which star formation could occur:
\begin{equation}
M_{min}^{halo} \approx 1.54 \times 10^5 M_\odot \Big(\frac{1+z}{31}\Big)^{-2.074}.
\label{eq:Hcooling}
\end{equation}
We take the maximum halo mass for Population III.1 star formation to be $M_{max}^{halo}=10^7 M_\odot$, though, due to the hierarchical nature of structure formation, our results are not sensitive to this choice. 

At some time between the beginning of Population III.1 star formation and the end of reionization at $z \sim 6$,
 massive Population III.1 (and Dark Star) formation must have given way to subsequent formation of less-massive stars~\cite{greifbromm2006},
 however there are few constraints on when this transition occurred.  Here, motivated by the work in Ref.~\cite{greifbromm2006}, we consider two 
scenarios for the termination of Population III.1 star formation; at redshifts $z_f=15$ and 11.
For each case, we assume that Population III.1 star formation was possible in any minihalo with a mass between $M_{min}^{halo}$ and $M_{max}^{halo}$ at redshift $z \geq z_f$.
 
It is unknown how many minihalos meeting the above criteria actually hosted Population III.1 stars.  We therefore parametrize 
the fraction that did as $f_s^0$.  The comoving number density of dark matter spikes as a function of redshift is then
\begin{equation}
N_{sp}(z) = f_{s}^0\left(1-f_{merged}(f_s^0,M_{BH})\right) N_{halo}(z),
\end{equation}
where $N_{halo}(z)$ is the comoving number density of minihalos in which Population III.1 star formation was possible, and $f_{merged}$, itself a function of $f_s^0$ and the black hole mass $M_{BH}$, is the fraction of dark matter spikes to have been destroyed in black hole mergers. In SDFS, we estimate that $f_{merged}$ is at most $1/2$, and is only significant for the largest black holes and $f_s^0 \sim 1$. For small $f_s^0$, $f_{merged} \rightarrow 0$ and the number density of dark matter spikes is determined only by $f_s^0$. In the following analysis, we parametrize the fraction of seemingly-capable minihalos to host a Population III.1 star and survive to become part of the Milky Way substructure today as $f_s$.

Assuming $f_s=1$, i.e.~that each viable minihalo hosted a Population III.1 star and that $f_{merged}$ is negligible, 
the $z=0$ distribution of dark matter spikes throughout the Galactic halo is obtained from the Via Lactea II (VL-II) cosmological N-body simulation~\cite{vl2} (see SDFS~\cite{dmspikes,dmspikesConf} for details).
We note that VL-II does not include the role of baryons in formation of the Galaxy (at epochs much
later than the initial formation of the first stars and their resultant black holes)  which could affect the numbers of spikes, most likely by contracting the dark matter halos and therefore also the spike distribution. However, the extent of this contraction, if it exists at all (e.g. the dark matter halo may actually expand during galaxy formation~\cite{Dutton:2006vi}), is uncertain.  We also note that dark matter spikes may be destroyed when two minihalos (each containing a black hole) merge, forming a close binary.  While this process is not included in our model, Ref.~\cite{dmspikes} includes an estimate of this effect: Mergers would change the number of dark matter spikes in our Galaxy today by at most a factor of 2 for the most massive black holes considered, and very little for low mass black holes.  Finally, dark matter subhalos near the Galactic center may experience tidal disruption, but the extent to which this affects dark matter spikes surrounding black holes is uncertain~\cite{bzs}.  In Ref.~\cite{dmspikes}, it was found that constraints on $f_s$ are generally robust with respect to uncertainties in the distribution of spikes near the Galactic center.  We return to this point in section~\ref{sec:fDS}.

\begin{figure}[h]
\begin{center}
\mbox{\epsfig{file=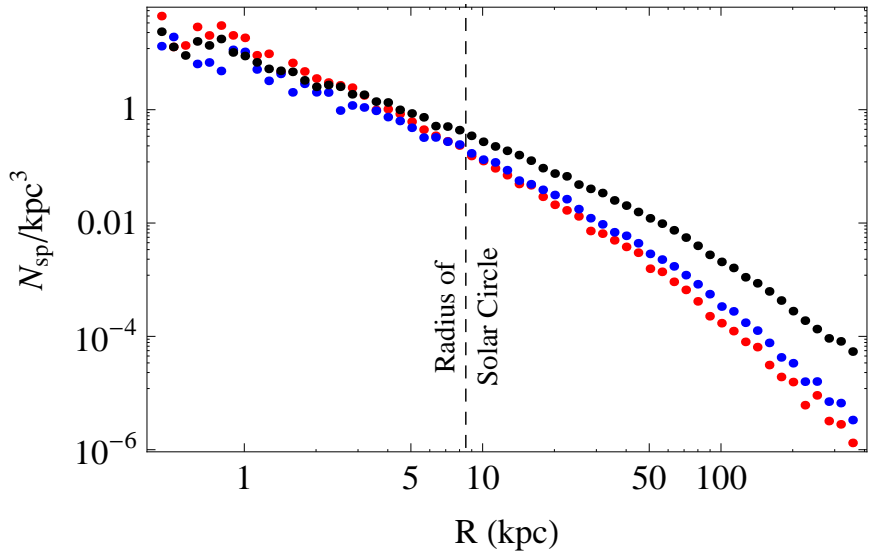,width=.7\textwidth}}
\end{center}
\caption{The number density of black hole spikes in the Milky Way as a function of Galactic radius for star formation models with $z_f=15$ (red) and $z_f=11$ (blue) as described in the text. We assume $f_s=1$.  The black points illustrate the total dark matter density profile at $z=0$.
\label{fig:spikedists}}
\end{figure}

In Fig.~\ref{fig:spikedists} we show the number densities of dark matter spikes inside the Milky Way halo as functions of Galactic radius for $z_f = 15$ (red) and$z_f = 11$ (blue). 
For larger $z_f$, Population III.1 star formation terminates earlier, so there were fewer stars, and therefore the fewer black holes and surviving density spikes today; we find 7983 spikes in our Galactic halo for $z_f=15$ and 12416 spikes in our Galactic halo for $z_f=11$.
In a similar analysis, Ref.~\cite{bzs} found $\sim1027$ spikes in our Galactic halo\footnote{As discussed in SDFS, our approach for finding relevant dark matter minihalos can be contrasted with that of Ref.~\cite{bzs}.
At $z=18$, they populated halos that constituted 3$\sigma$ peaks in the smoothed primordial density field with seed black holes of initial mass 100 $M_\odot$. Using an analytical model of halo evolution, they simulated 200 statistical realizations of the growth of a Milky Way-sized halo. Instead, we use one Galaxy-mass halo from a very high resolution cosmological simulation (VL-II) 
to follow potential Population III.1 star-forming minihalos to $z=0$. See~\cite{dmspikes} for further details.}. 
 For comparison, the total dark matter density profile at $z=0$ in VL-II is also shown (black);
 although the normalization of these points is
arbitrary, it is useful to illustrate that the total dark matter profile is more extended than the distribution of black holes with dark matter spikes.

The large population of dark matter spikes in our galaxy is consistent with the existence of high redshift quasars.  The Sloan Digital Sky Survey has identified high redshift quasars at $z\simeq6$, each presumably powered by a super massive black hole (SMBH), with a mass on the order of $10^9$ M$_\odot$.  These SMBHs grew from seed black holes~\cite{Haiman:2000ky}:  In the standard picture, the seed black holes formed from standard Population III stars, dark stars, or direct collapse models (See~\cite{Begelman:2007je} and references therein). 

Simulations indicate that most seed black holes grew by accreting baryons (see \cite{Bellovary:2010jr} and \cite{Bellovary:2011jq} for recent simulations), a scenario consistent with the formation of dark matter spikes. In fact, most seed black holes failed to become SMBH and end up as intermediate mass black holes (IMBH), with masses in the range $10^3$ M$_\odot$ to $10^5$ M$_\odot$. Simulations also indicate that a large fraction of the seed black holes only grew by accretion and never merged. Hence, dark matter spikes may well be ubiquitous.

The density profile of an individual dark matter spike surrounding a black hole is a key issue
in computing the annihilation signals.
 In this paper we make the best simple estimates we
 can and secondly demonstrate what we consider to be the worst case scenario.  
 The progenitors of the black holes discussed in this paper formed initially in the following way:
as a protostellar cloud started to collapse at the center of the dark matter halo, dark matter
was gravitationally pulled into the collapsing object along with the baryons.  The initial
density profile at the center of the halo then steepened due to this growing gravitational potential.  
We treat this gravitational dark matter enhancement with adiabatic contraction of the dark matter halo around the central mass.  A very simple prescription for the adiabatic contraction is the Blumenthal {\it et al.}~method~\cite{blum}, which only considers halo particles on circular orbits; equivalently, only their angular momentum is conserved as the halo is compressed.  We use this prescription in our simple 
estimates. However, early work on dark stars met with some skepticism because of 
exactly this issue.  Consequently several groups have verified that this simple Blumenthal {\it et al.}~method
is accurate to within a factor of two for the case of dark stars.  Iocco {\it et al.}~\cite{Iocco:2008rb}
(see Figure 1 in their paper)
used the Gnedin {\it et al.}~method~\cite{Gnedin:2004cx} and found basic agreement with the enhanced dark matter profiles obtained for these early stars using the
Blumenthal {\it et al.}~method.  Similarly, Natarajan, O'Shea, and Tan \cite{Natarajan:2008db} also used the Gnedin {\it et al.}~method and agreed.  
To address these concerns yet more carefully for the case of early stars forming at the centers
of minihalos, Ref.~\cite{Freese:2008hb} performed
an exact calculation using the Young method~\cite{young}, which takes into account the conservation of the radial action as well as angular momentum (the adiabatic invariants) and again reproduced the basic results found using the Blumenthal {\it et al.}~method (in some
cases finding densities higher by a factor of two and in others lower by a factor of two; 
see Figures 3 and 4 in their paper.)\footnote{It is important to point out that YoungÕs adiabatic prescription is not at all sensitive to departures from sphericity: Sellwood and McGaugh~\cite{sellwoodmcgaugh} showed that, for what concerns adiabatic contraction, even a flat disk could be well approximated as a sphere.}. Since
there are many uncertainties and parameter choices here (e.g.~the concentration
parameter, the redshift  of the object at the time of its formation, the WIMP mass (which
determines the baryon density at the time of dark star formation), the cutoff radius inside which we should assume the dark matter has already annihilated away by the time the black hole forms (see
below), the size it grows to, etc.), we cannot be more precise than this factor of two.  Hence we present two
sets of results:  First, we use
the Blumenthal {\it et al.}~method to obtain constraints.  Second, we present a worst case scenario,
namely, we obtain results under the assumption that spike densities are
lower by a factor of two relative to the Blumenthal {\it et al.}~prediction.  The latter analysis results in lower annihilation signals and far weaker bounds, and in this sense these are the most conservative results.

A second major source of uncertainty in the dark matter profile is
the choice of initial dark matter density profile prior to adiabatic contraction.
As our canonical case, we begin with Navarro, Frenk, and White (NFW) profiles~\cite{nfw} for both the baryons and dark matter, and, as described in the previous paragraph, use the Blumenthal {\it et al.}~prescription for adiabatic contraction~\cite{blum}.   However, it is possible that such an NFW
profile overestimates the amount of dark matter at the core of a dark matter minihalo where a Population III.1 star forms\footnote{We do note, however, that it is not unreasonable to imagine that
in these pristine early halos the central profile was steep while subsequent baryonic processes lowered the dark matter central cusp by the present epoch where observations are made.}.
To address this concern, Ref.~\cite{Freese:2008hb} looked at the extreme case of an
initial halo profile that has a core of constant dark matter density, known as a Burkert profile~\cite{burkert}. 
We note that this profile is not realistic for early minihalos and invoke it only as a scenario that results in the absolute minimum core dark matter density.
Even in this extreme case, the dark matter density at the core of the star forming in the center of the minihalo is only lower by about a factor of two.  We note that an initial Einasto profile~\cite{einasto} (in lieu of NFW
or cored) would result in a spike density in between these two extremes.  Thus we feel confident that
we may use the NFW case as a good estimate and present results as well for the worst
case where the dark matter density is lower by a factor of two due to the different possible initial profiles.

In short, to obtain the dark matter density profile around the black hole we use the Blumenthal {\it et al.}~method to obtain
simple estimates of adiabatic contraction starting from an initial NFW halo.  We then also
present results with densities lower by a factor of four:  a factor of two due
to the initial dark matter profile of the minihalo and another factor of two due to the success of adiabatic
contraction.  This factor of four represents our sense of the most extreme possible uncertainties in our results.  The sensitivity of our results to these choices is discussed in section~\ref{sec:fDS}, as well.

Finally, after adiabatic contraction, we impose a high density cut-off at small radii in the spike to account for the annihilation of the dark matter throughout the lifetime of the spike; 
\beq
\rho_{max} = \frac{m_{\chi}}{\langle \sigma v \rangle t_{BH}},
\label{eq:rhomax}
\eeq
where $m_\chi$ is the mass of the dark matter particle, $\langle \sigma v \rangle$ its annihilation cross section times velocity, and
$t_{BH}$ is the lifetime of the central mass, roughly $1.3 \times 10^{10}$ years for a star that formed at $z=15$.  

While standard Population III.1 stars are expected to have masses of $\sim 100 M_\odot$, if dark matter is
 capable of self-annihilating, then the first stars may have been powered by dark matter annihilations for some period of time prior to nuclear fusion.  This first phase of stellar evolution is known as the Dark Star phase~\cite{spolyar08} and may have lasted anywhere from a few hundred thousand years to millions, or even billions, of years.  During the Dark Star phase, the star remains cool enough to continue to accrete baryonic matter, and may grow to $\sim 1000M_\odot$~\cite{Freese:2008wh} or even as large as $\gtrsim 10^5 M_\odot$~\cite{Freese:2010re}, depending on the details of how the dark matter is depleted and replenished in the star.  We note that the WIMP mass does affect the final mass of the star, as the dark matter heating is inversely proportional to the WIMP mass, however the difference in the final stellar mass is less than a factor of two for 100 GeV and 1 GeV WIMPs assuming they move on circular orbits and dark matter is depleted in a simplistic way~\cite{Spolyar:2009nt}. If the dark matter particle orbits are more complicated, lighter WIMPs would result only in a cooler and more extended star during the Dark Star phase~\cite{Freese:2010re}.  

When the dark matter fuel inside a Dark Star runs out, it collapses and heats up to become a standard fusion-powered star, which, at the end of its life, will likely undergo core collapse leaving a black hole remnant. We use the potential existence of the Dark Star phase, as well as the possibility of direct collapse of very massive gas clouds to black holes~\cite{shapiro2004,Begelman:2007je}, to motivate consideration of black holes as large as $10^5 M_\odot$.


\section{Gamma Ray Signal from Dark Matter Annihilations}
\label{sec:signal}

For a Majorana dark matter particle with mass $m_\chi$ and annihilation cross section times velocity $\langle \sigma v \rangle$, the rate of WIMP annihilations in a dark matter spike is
\begin{equation}
\Gamma = \frac{\langle \sigma v \rangle}{2 m_\chi^2}\int_{r_{min}}^{r_{max}} dr \, 4 \pi r^2 \, \rho_{DM}^2(r),
\label{eq:rate}
\end{equation}
where $\rho_{DM}(r)$ is the dark matter density as a function of spike radius, with $r_{min}$ and $r_{max}$ defining the volume of the dark matter spike in which annihilations occur.

We choose as a benchmark scenario $\langle \sigma v \rangle = 3 \times 10^{-26}$ cm$^3$s$^{-1}$, in agreement with the measured dark matter abundance today for thermal WIMP dark matter, and consider several WIMP candidates, defined by mass and annihilation channel. We consider annihilations of WIMPs with masses of 5, 10, and 20 GeV 
to Standard Model final states $\tau^+ \tau^-$, $\mu^+ \mu^-$, and $d \bar d$.  Note that these WIMPs are not heavy enough to annihilate to $W^+W^-$. While
all but the lightest WIMPs considered here are heavy enough to annihilate to $b \bar b$, the resulting photon spectra from annihilation to all quark pairs ($q\bar{q}$) are nearly identical. We compute the spectrum of photons from annihilation to final state $f$, $dN_f/dE$, with PYTHIA \cite{pythia}, except for the final state $\mu^+\mu^-$, in which case the photon spectrum comes only from final state radiation~\cite{fsr}.
In the following analysis, we assume that the branching fraction to each final state is 1, though, in principle, some combination of final states is possible.


The differential flux of neutral particles from annihilations to final state $f$ in a dark matter spike with radius $r_{max}$ located some distance $D$ from our Solar System is given by
\begin{equation}
\frac{d\Phi_f}{dE}=\frac{\Gamma}{4 \pi D^2} \frac{dN_f}{dE},
\end{equation}
for $D \gg r_{max}$.  If $D$ is not much larger than $r_{max}$, and the spike is not point-like, an integral along the line-of-sight to the spike must be performed.  

If a single spike is a sufficiently bright and compact source of gamma-rays, it may have been identified as a point source and recorded in the FGST First Source Catalog~\cite{fgstFSC}, or, indeed, in the EGRET Source Catalog~\cite{EGRETsources}.
Spikes bright enough to be identified as point sources must not be brighter than the brightest source in the the FGST catalog.  Requiring that the flux not exceed this brightness establishes a minimal distance, $D_{min}^{PS}$, beyond which the spike must be located. Similarly, we can define a maximal distance, $D_{max}^{PS}$, as the distance beyond which a spike would not be bright enough to have been detected at $5 \sigma$ significance (see~\cite{dmspikes} and ~\cite{buckleyhooper} for details). All spikes which could have been identified as point sources lie between $D_{min}^{PS}$ and $D_{max}^{PS}$.

\begin{figure}[h!]
\begin{center}
\mbox{\epsfig{file=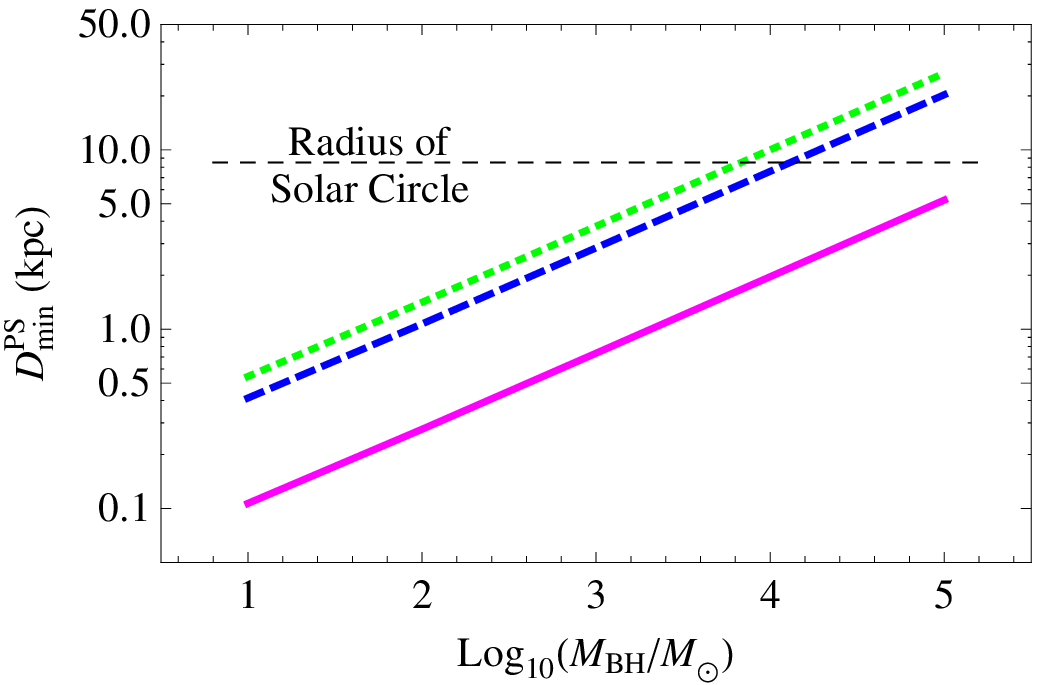,width=.48\textwidth}}\hspace{5mm}
\mbox{\epsfig{file=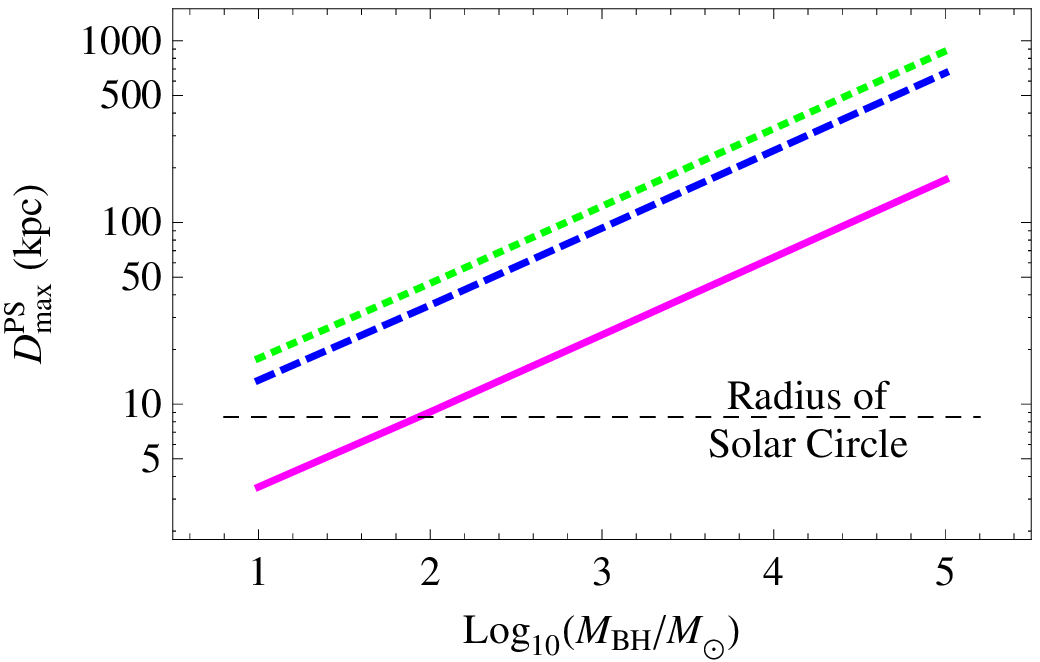,width=.48\textwidth}}
\caption{In the left panel, we display the minimal distance from our Solar System of a single spike such that it does not exceed the flux of the brightest source in the FGST First Source Catalog, $D_{min}^{PS}$, as a function of central black hole mass for Intermediate $z_f$.  From top to bottom, the contours are for 10 GeV WIMPs annihilating to $d \bar{d}$ (green dotted), $\tau^+\tau^-$ (blue dashed), and $\mu^+\mu^-$ (magenta solid).  In the right panel, we show the maximal distance from our Solar System of a single spike such that it would have appeared as a $\gtrsim 5\sigma$ point source to Fermi in the first year of data, $D_{max}^{PS}$, for the same cases. The horizontal black dashed line indicates the distance to the Galactic center from our Solar System.
\label{fig:PointSourceDist}}
\end{center}
\end{figure}

In the left panel of Fig.~\ref{fig:PointSourceDist}, we display
 the minimal distance, $D_{min}^{PS}$, at which a point source may be located in order not to exceed the largest flux from
 any point source measured by FGST for 10 GeV WIMPs annihilating to $d \bar{d}$ (green dotted), $\tau^+\tau^-$ (blue dashed), and $\mu^+\mu^-$ (magenta solid).  
For 10 GeV WIMPs annihilating to $d \bar{d}$ or $\tau^+\tau^-$, if the central black hole is $\lesssim 100 M_\odot$ (roughly what is expected from standard Population III.1 star formation), spikes may be located within 1 kpc of our Solar System. For annihilations to $\mu^+\mu^-$, spikes may be $\lesssim 1$ kpc from our Solar System even for black holes as large as $\mathcal{O}(10^3) M_\odot$.  However, for either choice of $z_f$ considered here, the VL-2 simulation results indicate less than one dark matter spike within 1 kpc of our Solar System, even for $f_s=1$.  Finally, we remind the reader that we consider only potential FGST point sources in this analysis; if dark matter is very light ($m_{\chi} \lesssim 5$ GeV) and annihilates to $\mu^+\mu^-$ in the spike surrounding a very small black hole ($M_{BH} \lesssim 10 M_\odot$), if the nearest spike is actually as near as $D_{min}^{PS}$ it may appear to be an extended gamma-ray source for FGST.  We do not consider extended sources here, though we return to this point in section~\ref{sec:fDS}.

The right panel of Fig.~\ref{fig:PointSourceDist} shows the maximal distance, $D_{max}^{PS}$, at which a single dark matter spike would have been identified by FGST as a $\gtrsim 5\sigma$ point source. Sources that are farther from our Solar System than $D_{max}^{PS}$ would be too faint to have been identified as point sources in the first year of FGST operations, and therefore the gamma-rays from these spikes would be part of the measured diffuse gamma-ray flux.  In fact, for 10 GeV WIMPs  annihilating to $\mu^+\mu^-$, if the typical black hole size is quite small ($\sim 10 M_\odot$), spikes within only a few kiloparsecs of our Solar System would be too faint to have been identified as gamma-ray point sources by FGST.  In all cases discussed here, for all but the largest black holes considered, at least some spikes in the Milky Way halo would contribute to the diffuse gamma-ray flux.


\section{Constraining $f_s$}
\label{sec:fDS}

As alluded to in section~\ref{sec:dmspikes}, it is quite possible that $f_s \ll 1$.  In this section we discuss how gamma-ray data can be used to constrain $f_s$, given a star formation scenario and a dark matter model.

First, it is possible to constrain $f_s$ with the measured diffuse gamma-ray flux:  Since the diffuse gamma-ray background is not yet well-understood, we conservatively require only that the total diffuse gamma-ray flux from dark matter annihilations around spikes in the Milky Way halo not exceed the measured flux in any of the nine FGST energy bins in Ref.~\cite{fgstEGB} by more than $3\sigma$. 
In SDFS, the diffuse gamma-ray flux measurements of FGST were found to constrain $f_s$ mainly for dark matter annihilating to leptonic final states for WIMP masses $\gtrsim 100$ GeV.
However, 
for the lighter WIMP masses $\lesssim20$ GeV considered in this paper, no interesting bounds result.  Because  the FGST measurements of diffuse flux indicate that it scales roughly as $E^{-2.4}$, diffuse gamma-ray constraints on dark matter annihilation tend to come from the highest energy bins, if they exist at all.  For light WIMPs (relative to standard ~100 GeV WIMPs), the entire gamma-ray spectrum is shifted to lower energies such that the gamma-ray flux predicted for WIMPs annihilating in dark matter spikes is not in conflict with FGST measurements.  
Hence we restrict our discussion in the remainder of the paper to point sources observed by FGST.

A second way to constrain $f_s$, making use of FGST point source data, is to require that the integrated luminosity of the brightest (i.e.~closest) dark matter spike not exceed the integrated luminosity of the brightest Fermi point source.
As discussed above, the left panel of Fig.~\ref{fig:PointSourceDist} displays this minimal distance, $D_{min}^{PS}$, for 10 GeV WIMPs.  A limit on $f_s$ may be established by requiring the expectation of finding less than one spike within a sphere of radius $r = D_{min}^{PS}$ centered on our Solar System.  
That is,
\beq
\int_0^{D_{min}^{PS}}r^2dr\int_0^{4 \pi}d\Omega \, N_{sp}(R,f_s) \leq 1,
\eeq
where $N_{sp}(R, f_s)$ is the number density of spikes as a function of Galactic radius, $R$, and the fraction of spikes to survive, $f_s$.  As discussed in section~\ref{sec:dmspikes}, we assume that the actual number density of spikes, $N_{sp}(R,f_s)$, is related to the total possible number density of spikes, $N_{sp}(R,f_s=1)$, as
\beq
N_{sp}(R,f_s) = f_s \times N_{sp}(R,f_s=1).
\eeq

\begin{figure}[h!]
\begin{center}
\mbox{\epsfig{file=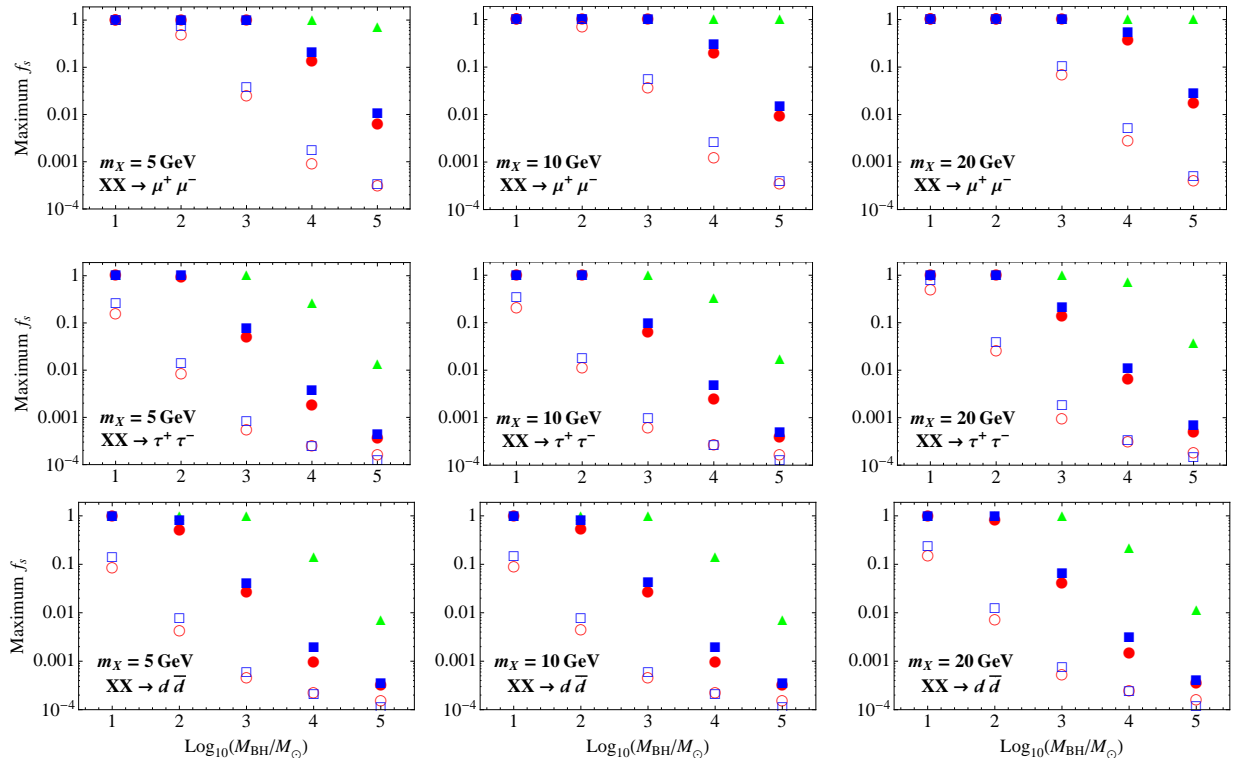,width=1.0\textwidth}}
\end{center}
\caption{\it Upper limits on the fraction $f_s$ of minihalos to host the first stars, plotted as a function of the mass of the remnant black hole at the center of each dark matter spike.  These limits are derived from the gamma ray flux from point sources measured by FGST.  Squares and circles
correspond to our best estimates while the green triangles illustrate the extreme case of most
conservative bounds as described in the text.
The filled points are the upper limits obtained by requiring the dark matter annihilation signal in gamma-rays not exceed the 
 brightness of Vela, while the open points illustrate the upper limits obtained by requiring the dark matter annihilation gamma-ray flux not exceed that from the brightest unassociated FGST point source. Intermediate and Late $z_f$ are shown as red circles and blue squares, respectively.
\label{fig:fDSmaxp}}
\end{figure}

In Figure~\ref{fig:fDSmaxp}, we plot the maximal allowed value of $f_s$ as a function of black hole mass for $z_f=15$ (red circles) and $z_f=11$ (blue squares).
 The filled points correspond to the constraint derived from the requirement that the brightest spike not be brighter than the brightest FGST source (corresponding to $D_{min}^{PS}$ as plotted in Fig.~\ref{fig:PointSourceDist}), which is known to be associated with the Vela pulsar. Although we cannot exclude the possibility that the brightest dark matter spike is located along our line of sight to Vela, it would be a bizarre and grand conspiracy if all of the brightest dark matter spikes were to be located along our lines of sight to the brightest gamma-ray point sources, many of which are associated with known astrophysical objects. Thus, alternatively, one could consider the constraints derived from the requirement that the brightest dark matter spike not be brighter than the brightest {\it unassociated} FGST point source; that is, a point source that is not associated with any known astrophysical object. These less conservative, but perhaps more reasonable, constraints are shown as the open points in Figure~\ref{fig:fDSmaxp}.  

We also show, as solid green triangles in each panel of Figure~\ref{fig:fDSmaxp}, the constraints resulting from a dark matter density spike that is less dense by a factor of four as discussed above 
Equation~\ref{eq:rhomax} as the most extreme possibility; for these points, we 
 required that the brightest spike not be brighter than the brightest FGST source and show only the star formation history leading to the weakest constraints.  They therefore represent the most conservative limits on $f_{s}$, accounting for uncertainty in the contracted halo profile.  These limits (solid green triangles) represent the ``worst case scenario" for indirect detection (among those considered here).
 We also note that the difference between the solid and open blue and red points is a factor of 4.7 in spike distance or, equivalently, spike density,
 so that the green triangles for the case of brightest unassociated FGST point source would
 lie slightly below the solid blue points (again, assuming the less favorable scenario for the termination of Population III.1 star formation).
 We remind the reader that throughout we have assumed a thermal annihilation cross section, $\langle \sigma v \rangle = 3 \times 10^{-26}$ cm$^3$s$^{-1}$; a larger (smaller) annihilation cross section would lead to stronger (weaker) constraints than those shown in Figure~\ref{fig:fDSmaxp}.

In all cases, constraints are strongest for the largest black holes, which result in the most dense spikes, and for the lightest dark matter, for which the number density in a spike is the largest. For $\chi \chi \rightarrow \mu^+ \mu^-$, shown in the top panels, the gamma-rays come exclusively from final state radiation, resulting in relatively weak constraints.  For example, if the typical black hole mass is ${\cal O}(10-100) M_\odot$, the fraction $f_s$ is unconstrained for any dark matter mass $\gtrsim 5$ GeV. However, even for such light dark matter, the constraints can be quite strong if the typical black hole mass is larger. In the middle and bottom panels, for $\chi \chi \rightarrow \tau^+ \tau^-$ and $\chi \chi \rightarrow d \bar{d}$, respectively, annihilations lead to states that either hadronize or decay to states that hadronize, resulting in a larger gamma-ray flux and therefore stronger constraints on $f_s$. Remarkably, in these cases constraints are at the percent level for black holes of mass $\sim 100 M_\odot$, the typical mass expected from standard Population III.1 star formation (in the absence of a Dark Star phase).

As discussed in section~\ref{sec:signal}, we consider only FGST point sources in this analysis.  If the dark matter is very light ($m_{\chi} \lesssim 5$ GeV) and annihilates to $\mu^+\mu^-$ in the spike surrounding a very small black hole ($M_{BH} \lesssim 10 M_\odot$), and if the nearest spike is actually as near as $D_{min}^{PS}$, it may appear to be an extended gamma-ray source for FGST.  In the upper left panel of Fig.~\ref{fig:fDSmaxp}, it is apparent that any scenario in which there is the possibility that the nearest spike might appear as an extended source is entirely unconstrained anyway.  An analysis of extended sources may prove very interesting: it may be possible to discover the distinct shape of a dark matter spike in this manner, though no additional constraints would result for the scenarios considered here.

In principle it might be prudent to exclude the inner few kpc from the Galactic center from our study because spikes in the inner region of the Galaxy may have been disrupted, though
in practice the results do not differ significantly from what we report here.  The origin of the insensitivity to the Galactic center region can be understood by considering the following:
First, models in which $D_{min}^{PS} \lesssim$ 3.5 kpc are not affected at all by what happens within $\sim 5$ kpc of the Galactic center.  Second, for models in which $D_{min}^{PS}$ is comparable to the radius of the Solar Circle, we demand less than one spike within a volume of $\sim 10^3$ kpc$^3$ centered on our Solar System, already orders of magnitude below the expected number density of spikes in the Solar neighborhood, so these cases are strongly constrained whether or not the spikes nearest the Galactic center are considered.  Furthermore, the vast majority of spikes are not located near the Galactic center; for $z_f=15$ and 11, the percentages of Milky Way spikes within 5 kpc of the Galactic center are just 10\% and 4\%, respectively.  Thus, the constraints on $f_s$ are robust with respect to uncertainties in the dynamics near the Galactic center.

Finally, we would like to stress that it is possible that one or more point sources in the FGST catalog are, in fact, dark matter spikes.  In this case, spectral information may provide some clues as to the nature of dark matter.  While a spectral analysis is beyond the scope of this study, Ref.~\cite{buckleyhooper} finds that as many as 20 to 60 gamma-ray point sources may in fact be dark matter substructure. Perhaps we have more information about dark matter spikes than we imagine.

\section{Discussion and Conclusion}

We have calculated the limits on the fraction of viable dark matter minihalos in the early universe to host Population III.1 stars, surviving today as dark matter spikes in our Milky Way halo.  Motivated by potential hints of light dark matter from the DAMA and CoGeNT direct dark matter searches, we considered thermal relic WIMP dark matter with masses of 5, 10, and 20 GeV, and annihilation to $\mu^+\mu^-$, $\tau^+\tau^-$, and $q\bar{q}$.  We find that constraints from the integrated luminosity of Vela are only significant if the typical black hole mass is $\gtrsim 10^3 M_\odot$, as would be expected if the first stars experience a Dark Star phase of stellar evolution, during which the star is powered by dark matter annihilations.  Excluding the very unlikely coincidence that all of the brightest dark matter spikes are located along our lines of sight to the brightest known astrophysical gamma-ray sources, then the constraints on $f_s$ are at the percent level even for $100 M_\odot$ black holes if dark matter annihilates primarily to quark pairs or tauons.  Even in our ``worst case scenario" for indirect detection of dark matter spikes, constraints are significant for typical black hole masses $\gtrsim 10^3 M_\odot$.

From this brief study we conclude that, if dark matter is light, either the typical black hole size is $\lesssim 100M_\odot$ (i.e.~ there is no significant Dark Star phase), and/or dark matter annihilates primarily to $\mu^+\mu^-$ or other final states that result in low gamma-ray luminosity, and/or that an extremely small fraction of minihalos in the early universe that seem to be suitable to host the formation of the first stars actually did.  We look forward to additional evidence of early star formation and the properties of dark matter, and, eventually, to a solid understanding of both.

\section*{Acknowledgements}

J.D.~is supported by the Swiss National Science Foundation.
K.F.~thanks the Department of Energy and the Michigan Center for Theoretical Physics for support, and the Aspen Center for Physics for hospitality during the course of this research. 
KF thanks the Texas Cosmology Center (TCC) where she is a Distinguished Visiting Professor. TCC is supported by the College of Natural Sciences and the Department of Astronomy at the University of Texas at Austin and the McDonald Observatory.
K.F.~thanks Paul Shapiro for helpful conversations.
P.S.~is supported by the National Science Foundation under Grant No.~PHY-0969020, and by the University of Utah. 
D.S.~is supported by the Department of Energy.


\end{document}